# Panel-Scale Reconfigurable Photonic Interconnects for Scalable AI Computation

TZU-CHIEN HSUEH (Senior Member, IEEE), BILL LIN (Senior Member, IEEE), ZIJUN CHEN (Graduate Student Member, IEEE), and YESHAIAHU FAINMAN (Life Fellow, IEEE)

Department of Electrical and Computer Engineering, University of California, San Diego, La Jolla, CA 92093 USA

CORRESPONDING AUTHOR: Tzu-Chien Hsueh (e-mail: tzhsueh@ucsd.edu).

This work was supported by the National Science Foundation under Award 2023730, Award 2025752, Award 2045935, Award 2217453, and Award 2410053; and in part by ASML/Cymer Corporation.

**ABSTRACT** Panel-scale reconfigurable photonic interconnects on a glass substrate up to 500-mm × 500-mm or larger are envisioned by proposing a novel photonic switch fabric that enables all directional panel-edge-to-panel-edge reach without the need for active repeaters while offering high communication bandwidth, planar-direction reconfigurability, low energy consumption, and compelling data bandwidth density for heterogeneous integration of an in-package AI computing system on a single glass-substrate photonic interposer exceeding thousands of centimeters square. The proposed approach focuses on reconfigurable photonic interconnects, which are integration-compatible with commercial processor chiplets and 3D high-bandwidth memory (HBM) stacks on a large-area glass substrate, to create a novel panel-scale heterogeneously integrated interposer or package enabling low-energy and high-capacity wavelength-division-multiplexing (WDM) optical data links using advanced high-speed optical modulators, broadband photodetectors, novel optical crossbar switches with multi-layer waveguides, and in-package frequency comb sources.

## I. INTRODUCTION

The exponential rise in the complexity of artificial intelligence (AI) and machine learning (ML) models continues to explode at a rapid pace. For example, the well-known ChatGPT [1] large language model (LLM) was initially developed in 2018 as GPT-1, with 117 million parameters, which quickly grew by more than ten times to 1.5 billion parameters with GPT-2 in 2019. It then grew by another 100 times to 175 billion parameters with GPT-3 in 2020 [2]. The current GPT-4 model was released in 2023 with an estimated 1.7 trillion parameters [2], growing the model by yet another 10 times in comparison with GPT-3. Remarkably, in just five years, the LLM underlying ChatGPT has increased in size by over 10,000 times. The next generation of LLMs is expected to grow by yet another order of magnitude or two, with the corresponding computing requirements for training and inference growing at least a similar rate. This super-exponential growth is likely to continue, as there appears to be no scaling limitations in terms of learning capacity with respect to the underlying LLM architecture [3] or the availability of data from which to learn. Similar explosive growth is occurring in other AI and ML models, particularly with generative AI (GenAI) applications, such as text-to-image and text-to-video generation [4]. At the same time, there is an exponential rise in AI users and their frequency of use, creating another dimension of explosion in computing demands for run-time inferences [5]. These different dimensions of exponential growth mean that existing computing architecture approaches will surely be unable to keep up with the disruptive increases in computational and energy requirements. Overall, the size of GenAI models has been exploding at a rapid pace, increasing by more than an order of magnitude every year for the past five years.

With Moore's law slowing down, the ability to increase the computing density on a single die is diminishing, far from keeping up with the exponential growth in computing demand for GenAI. Emerging heterogeneous integration technologies enable the continued scaling of compute density to meet growing compute demand by integrating multiple processor chiplets (XPUs) with 3D stacks of high-bandwidth memories (HBMs) into a single advanced silicon interposer or package. However, enabling communication between the 3D-HBM stacks and processor chiplets is challenging [6] because the electronic interconnects on existing silicon interposers are limited to one link per physical interconnection as well as to a few centimeters per link distance at high data rates due to signal crosstalk and losses.

In this paper, panel-scale reconfigurable photonic interconnects on a glass substrate up to 500-mm × 500-mm or larger are envisioned by proposing a novel photonic switch fabric that enables arbitrary panel-edge-to-panel-edge reach without the need for active repeaters while offering high communication bandwidth, planar-direction reconfigurability, low energy consumption, and compelling data bandwidth density for heterogeneous integration of an in-package AI computing system on a single glass-substrate photonic interposer exceeding thousands of centimeters square. As the scalability of this photonic interposer continues to advance, it is expected that more than 100 Tb/s of data bandwidth will be transmitted and received in and out of an interposer carrying the entire in-package AI computing system [7], [8]. In short, the proposed approach focuses on reconfigurable photonic interconnects, which are integration-compatible with commercial processor chiplets and 3D-HBM stacks on a large-area glass substrate, to create a novel panel-scale heterogeneous integrated interposer or package enabling low-energy and high-capacity wavelength-division-multiplexing (WDM) optical data links using advanced high-speed optical modulators, broadband photodetectors (PD), novel optical crossbar switches (X-bar SW) with multi-layer waveguides, and in-package frequency comb sources.

The remainder of the paper is organized as follows. Two types of glass-substrate photonic interposers are proposed in Section II. The major features and comparisons between today's silicon interposers and our proposed glass-substrate photonic interposers are analyzed and summarized in Section III. The key building blocks under development for the proposed glass-substrate photonic interposers are reported and discussed in Section IV. The vision, future potential, and development roadmap are summarized in Section V.

## II. GLASS-SUBSTRATE PHOTOIC INTERPOSERS

Two types of scalable and reconfigurable glass-substrate photonic interposers are proposed: one is a photonic-integrated-chip (PIC) [9] embedded glass interposer (GIP), and the other is a monolithic-photonics (MPH) integrated glass interposer. Both PIC-embedded and MPH-integrated GIPs utilize a single glass substrate to support heterogeneous integration among electronic integrated chiplets (EIC), such as XPUs, controllers, 3D-HBMs, and accelerators, forming the entire in-package AI computing system. More importantly, both leverage photonic interconnections to enable high-speed, wide-bandwidth, high-connectivity, high-reconfigurability, and low-power in-package chiplet-to-chiplet communications among the EICs without compromising required heat dissipation and robust power delivery. Additionally, their architectural features, described in the following subsections, enhance the scalability of their in-package integrations with panel-scale packaging and wafer-scale manufacturing technologies, respectively.

### A. THE PIC-EMBEDDED GLASS INTERPOSER

The conceptual top view and zoomed-in cross-section view of the PIC-embedded GIP are shown in Fig. 1, where multiple fabricated PICs are embedded in a single piece of glass substrate through post-silicon integration, alignment, and assembly processes. This PIC-embedded GIP can be viewed as a short-term goal of the in-package photonic interconnect solution, as more than 95% of the fabrication, integration, and assembly technologies required for achieving this solution are commercially available.

The main idea is that all the EICs are 3D-assembled on the surface of the GIP through multiple embedded PICs, where the EICs are flipped chips and assembled by the downward micro-bumps of the upper EICs and the upward micro-pads of the lower PICs or EICs, as shown in Fig. 1(a). More importantly, all the PICs, highlighted in yellow, are post-silicon assembled into the pre-carved or pre-etched slots in the glass substrate, highlighted in light blue. Thus, these PICs play a key transition role between the 3D and 2D interconnects of the entire AI computing system integrated in this GIP. In the 3D direction, each PIC transits high-speed data from the EICs above through the PIC micro-pads, PIC metal/via stacks, PIC electrical-to-optical (E/O) or optical-to-electrical (O/E) conversions to the PIC optical waveguides (WG), as well as low-speed data/power deliveries between the EICs above and GIP below through PIC micro-pads, PIC metal/via stacks, PIC through-substrate vias (TSV), and PIC bumps. In the 2D directions, when the optical WGs reach the edges of each PIC, the PIC edge couplers and GIP fiber arrays can seamlessly bridge the high-speed optical interconnects among all the PICs, as shown in Fig. 1, with the proposed planar optical X-bar SWs in the PICs, elaborated in Section IV, to enable fully reconfigurable chiplet-to-chiplet communication between any two EICs integrated in this in-package AI computing system. Note that when assembling each PIC into its pre-carved slot in the glass substrate, the alignment between the PIC edge-coupler and GIP fiber arrays can be done by referencing the dummy edge couplers and fibers at the two ends of each array [10].

The primary functions of the glass substrate are described as follows: one is to complete internal high-speed optical communication network by stitching all the pieces of the PICs together through the GIP fiber arrays and the PIC edge couplers as well as external optical interfaces at the GIP outer edges, as shown in Fig. 1; the other is to provide a redistribution layer (RDL) for the 2.5D transitions of all the bump/pad pitches, electronic power supplies, biases, thermal controllers, and low-speed data communications between the internal and external of this PIC-embedded GIP containing the entire in-package AI computing system.

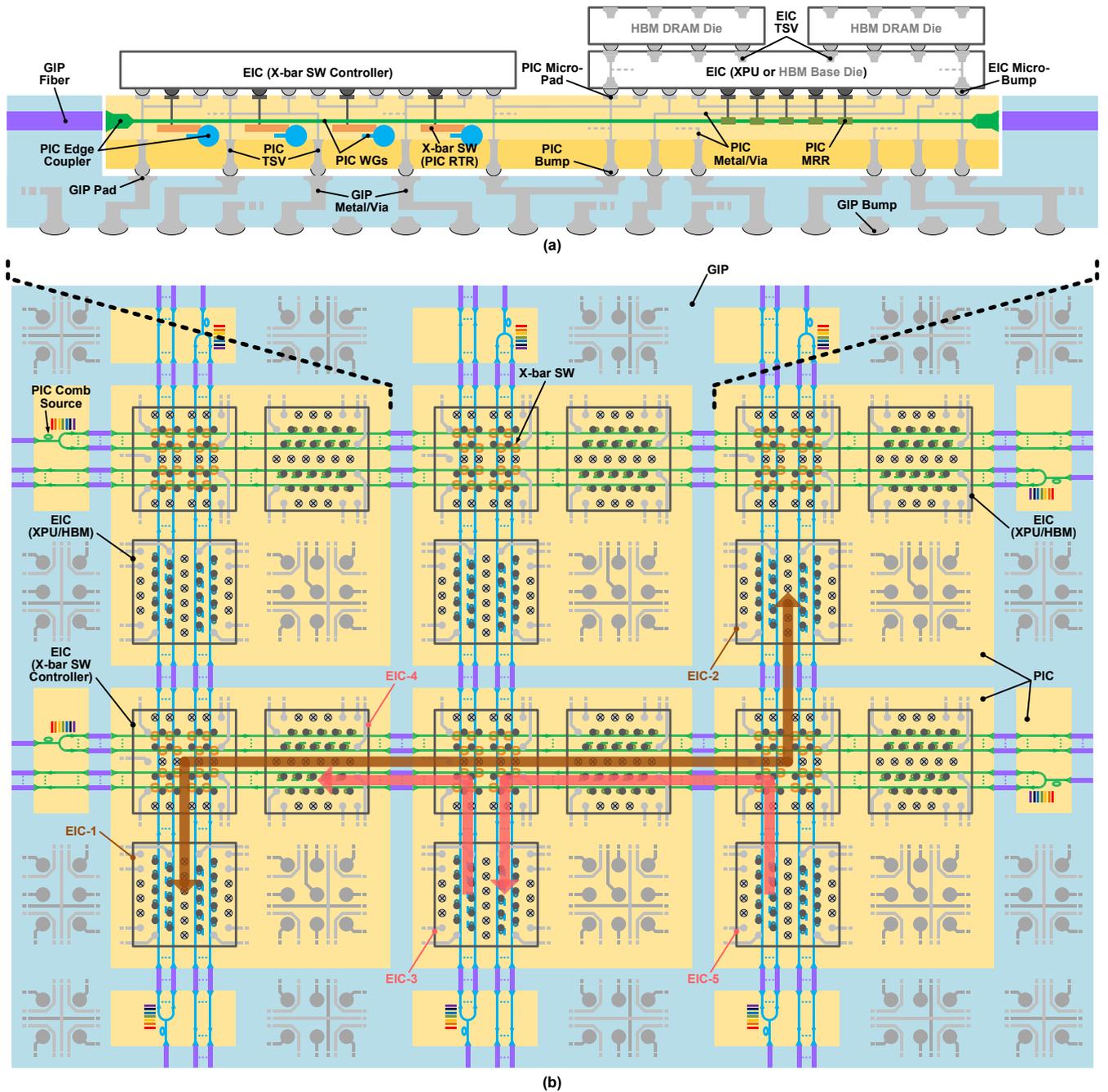

**FIGURE 1.** (a) The zoomed-in cross-section view of the PIC-embedded GIP. (b) The top view of the scalable and reconfigurable PIC-embedded GIP, which contains, for example, six photonic unit-interposers with their EICs and X-bar SWs, networking through broadband mesh-structured photonic interconnects, to form an in-package AI computing system.

In general, the EICs and PICs are individually fabricated by their own chip-level CMOS, memory-device, and silicon-photonics semiconductor manufacturing technologies. On the other hand, it is important to note that the glass substrate, which contains the fiber arrays and RDL, is produced using glass-panel manufacturing technologies. In other words, the dimension of this PIC-embedded GIP can be as large as a panel scale, which is determined by the panel sizes that can be handled by state-of-the-art glass-panel manufacturing and panel-level packaging (PLP) equipment and tooling to form the complete PIC-embedded GIP. For example, a standard range of today's glass-panel sizes is around 510-mm × 510-mm to 600-mm × 600-mm for the PLP [11], [12]. In short, the EICs and PICs are chip-level products, whereas the GIP can be a panel-level product, so the entire AI computing system integration can be accomplished through the panel-level heterogeneous assembly of the three components: EICs, PICs, and GIP.

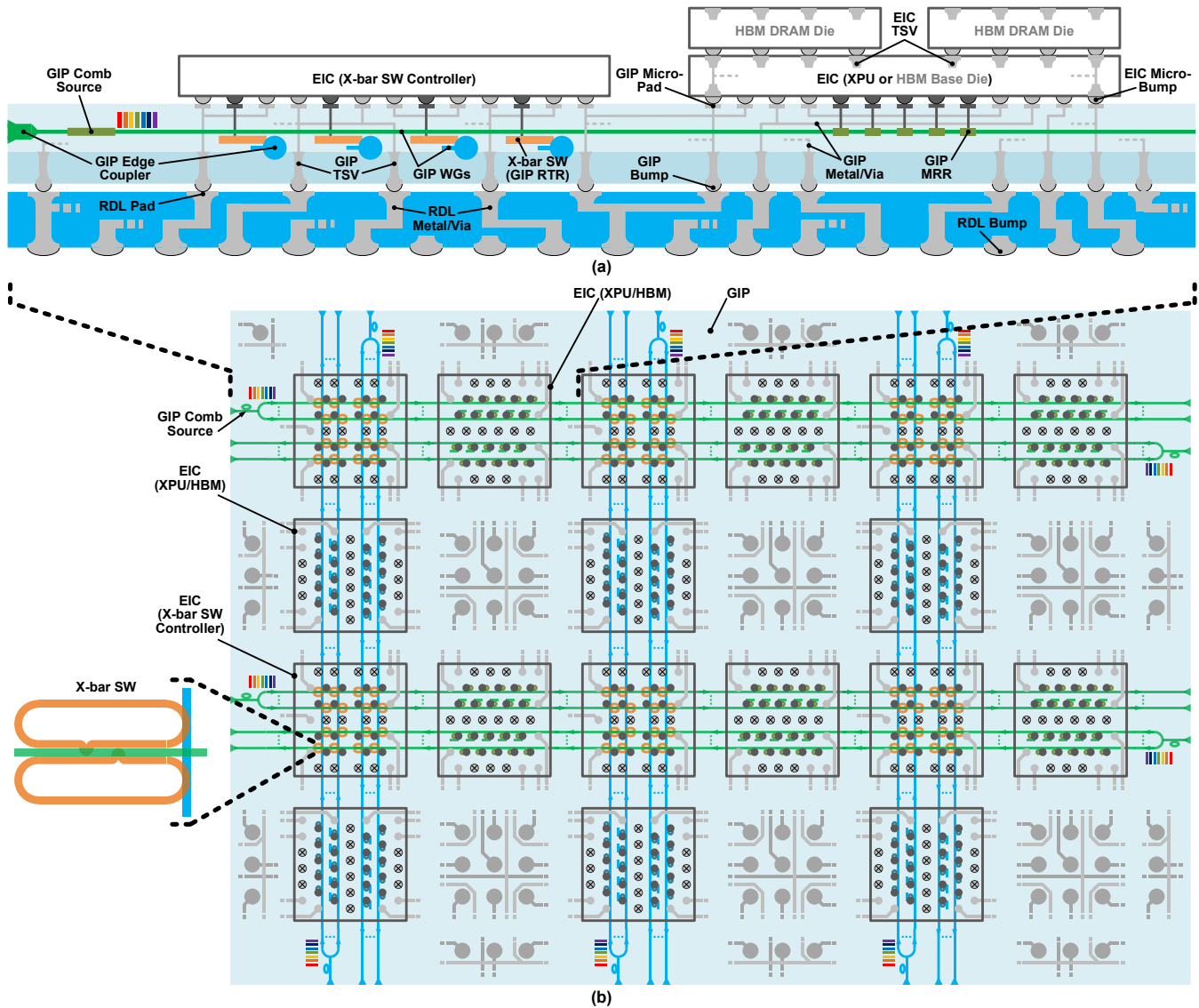

**FIGURE 2.** (a) The zoomed-in cross-section view of the MPH-integrated GIP. (b) The top view of the scalable and reconfigurable MPH-integrated GIP, which contains, for example, six photonic unit-interposers with their EICs and X-bar SWs, networking through broadband mesh-structured photonic interconnects, to form an in-package AI computing system.

### B. THE MPH-INTEGRATED GLASS INTERPOSER

The conceptual top view and zoomed-in cross-section view of the MPH-integrated GIP are shown in Fig. 2, where the photonic and electronic layers, including all the required photonic E/O and O/E devices, WGs, edge couplers for high-speed data communications as well as electronic metal/via stacks, TSVs, micro-bumps/pads for low-speed data and power deliveries, are monolithically fabricated on a single piece of glass substrate through the semiconductor manufacturing procedures, including implantation, deposition, planarization, doping, etching, and lithography processes. The RDL assembled beneath the GIP, as shown in Fig. 2(a), can be economically fabricated using a package-level or printed-circuit-board (PCB) manufacturing process.

It is important to note that although there are several pros and cons between silicon and glass substrates [13], the glass substrate used for this MPH-integrated GIP can be a modern silicon-on-glass (SOG) solution or one of its popular variants, which is referred to as a silicon-on-insulator (SOI) solution depositing silicon-photonics layers onto a glass substrate, as mentioned, to gain complementary benefits from both silicon and glass materials [14], [15], including better temperature tolerance, mechanical durability, optical bandwidth expansion, and dielectric breakdown tolerance. With all these features, the long-term goal is to fabricate this MPH-integrated GIP on a wafer scale [16], whose dimensions are determined by the wafer diameters that can be handled by state-of-the-art semiconductor manufacturing and wafer-level packaging (WLP) equipment and tooling [17], [18], [19]. For example, a maximum 300-mm × 300-mm glass substrate can be produced from the most advanced 450-mm-diameter glass wafers. In short, the EICs and MPH-integrated GIP are chip-level and wafer-level semiconductor products, respectively, while the RDL is a package-level product. The entire AI computing system can be integrated through the heterogeneous assembly of these three components: EICs, MPH-integrated GIP, and RDL, which is similar to today's silicon-interposer system integration approaches [20], [21], but on a wafer scale.

Although the EIC integration floor plans and photonic interconnect structures of the PIC-embedded and MPH-integrated GIPs shown in Fig. 1(b) and 2(b), respectively, are similar, the MPH-integrated GIP has about 45% better integration density and data communication density (bits/s/mm$^2$) since the advantage of the monolithic photonics fabrication on a glass substrate eliminates the chip boundary clearance areas for material transition and post-silicon assembly between the PIC and GIP in cases of the PIC-embedded GIP. However, the overall cost of semiconductor manufacturing and heterogeneous assembly for the MPH-integrated GIP could still be higher, primarily due to its wafer-scale fabrication per GIP and secondarily due to the assembly procedure required for attaching the RDL beneath a thin and large piece of the wafer-scale glass substrate, as shown in Fig. 2(a). Furthermore, the scalability of the AI computing system integrated into the MPH-integrated GIP may eventually be limited by the achievable wafer sizes, which have shown minimal progress over the past two decades [22]. On the other hand, panel-level manufacturing and packaging appear to be no critical bottleneck when scaling up to the next order of magnitude in area. Instead, the scalability of the AI computing system integrated into the PIC-embedded GIP may be dominated by the optical signal loss when the chiplet-to-chiplet communication signal bypasses too many optical X-bar SWs and edge couplers, which is further discussed in Section IV.

## III. ADVANCED FEATURES OF GLASS-SUBSTRATE PHOTOIC INTERPOSERS

Compared to today's silicon interposers, both the panel-scale PIC-embedded and wafer-scale MPH-integrated GIPs possess multiple advanced features, including high reconfigurability with 10× longer interconnect distances, high AI workload scalability, and better bandwidth density under certain energy efficiency and micro-bump usages, because of the unified optical data communication interface and repeatable planar mesh-structure floor plan for efficient EIC integration, photonic signal routing, and X-bar SW arrangements. For the sake of simplicity, the elaborations in the following subsections mainly use the PIC-embedded GIP as the examples, which are fully applicable to the cases of the MPH-integrated GIP.

### A. INTERCONNECT RECONFIGURABILITY

The assembly-efficient planar mesh structure of the GIP shown in Fig. 1(b) or 2(b) supports dynamic chiplet-to-chiplet communication among the EICs of the in-package AI computing system, as the proposed optical X-bar SWs at the WG crossing nodes (i.e., planar intersections between the blue and green WGs) are fully reconfigurable in a time-multiplexing manner. In other words, at a particular moment of time, any two specific EICs, e.g., an XPU and a nearby or far-off 3D-HBM, within this GIP can communicate with each other through programmable direction controls from the X-bar SW controllers. Although these two EICs are hard assembled at the designated locations of the GIP, which may potentially be several hundred millimeters apart from each other, this reconfigurability allows them to communicate with each other entirely in the photonics domain without intermediate O/E/O conversions or electronic buffering. For example, one option of the reconfigurable interconnect routes between EIC-1 and EIC-2 is shown in Fig 1(b), whose the worst-case scenario (i.e., highest optical loss) is not necessarily the longest distance between the two EICs but the maximal number of planar turns through the X-bar SWs, which is limited to maximum two 90-degree turns as the data path highlighted in brown; the insertion losses of the X-bar SW are discussed in Section IV. In addition, this photonic interconnect reconfigurability can also support chiplet-to-chiplet communication among three EICs, for example, the pink data paths in Fig. 1(b), where EIC-3 can simultaneously transmit data to EIC-4 and receive data from EIC-5, because the data transmitting paths from EIC-3 to EIC-4 are independent of the data receiving paths from EIC-5 to EIC-3, including electrical transmitters (TX), E/O modulators, WGs, X-bar SWs, O/E PDs, and electrical receivers (RX), as the unidirectional data links in most data communication standards.

All data path assignments and reconfigurability are controlled and governed by the X-bar SW controllers flip-chip assembled on top of the PIC areas containing the X-bar SWs and WG crossing nodes, as shown in Figs. 1 and 2, in a time-multiplexing manner. For example, the brown and pink data paths in Fig. 1(b) cannot be enabled simultaneously because of partial overlaps between their X-bar SWs and WGs. Additionally, when the brown data path is enabled, except for EIC-1 and EIC-2, any E/O and O/E interfaces of the other EICs attached to this data path should be disabled to prevent interrupting the data exchange or communication between EIC-1 and EIC-2. Overall, although any WGs, X-bar SWs, E/O interfaces, and O/E interfaces can only be configured to serve a particular data path between two certain EICs at a specific moment, this photonic interconnect reconfigurability can dramatically improve the data accessing efficiency by utilizing time gaps due to the idle time percentage of XPUs (i.e., 10% to 50%) without any data repeaters across a more than 500-mm distance.

To be specific, the comparisons of two scenarios using today's silicon interposer and the proposed photonic interposer to support an XPU nearby and far-off chiplet-to-chiplet communications are shown and summarized in Fig. 3 and Table 1, respectively. Both scenarios feature an XPU with a 768-mm$^2$ die area [23] and twelve 3D-HBM stacks with an 84-mm$^2$ base-die area [24] per stack. Therefore, the unit-interposer areas for both scenarios can be roughly identical at 3534 mm$^2$ (= 62 mm × 57 mm). Also, the comparison tends to equate the micro-bump usage for both scenarios to be 832 micro-bumps, as detailed below.

In the scenario of a silicon unit-interposer shown in Fig. 3(a), the XPU communicates with its nearby 3D-HBM stacks through the HBM PHY standard [25] with thirty-two 16-Gb/s single-ended data links [26], [27], [28] per 3D-HBM stack to support a total of 6.144 Tb/s nearby data bandwidth across twelve 3D-HBM stacks. Meanwhile, the far-off data communications between the XPU and XPUs or 3D-HBMs outside the 62-mm × 57-mm gray-boxed unit-interposer area are accomplished through a two-step data communication process. First, the XPU communicates with its nearby XSR SerDes dies through the UCIe PHY standard [29], [30], [31] with one hundred and twelve 32-Gb/s single-ended data links [32], [33] per XSR SerDes die to first transit a total of 7.168 Tb/s data bandwidth from the XPU to the nearby XSR SerDes dies. Second, the XSR SerDes dies buffer the entire 7.168 Tb/s far-off data bandwidth to other EICs on other silicon unit-interposers located at the north and south of the current silicon unit-interposer, as shown in Fig. 3(a), through the XSR SerDes PHY standard [34] with thirty-two 112-Gb/s differential data links [35], [36], [37], [38], [39], [40], [41] per XSR SerDes die, which can support up to 50 mm long electronic interconnect distances. Overall, the total data bandwidth of the central XPU is 13.312 Tb/s (= 6.144 Tb/s + 7.168 Tb/s) directly connected by the HBM and UCIe data links, which are all single-ended signaling inputs/outputs (I/O) with one 40-µm pitch micro-bump per link [21]. Therefore, a total of 832 (= 12×32 + 2×112 + 2×112) micro-bumps in the XPU and XSR SerDes dies are required for the nearby data links, a total of 128 (= 64×2) 110-µm pitch bumps are required in the XSR SerDes dies for the far-off data links [21], and a total of 736 (= 12×32 + 2×112 + 2×64) transmission lines (TL) are required for all single-ended nearby (i.e., HBM and UCIe) and differential (i.e., XSR SerDes) far-off data links.

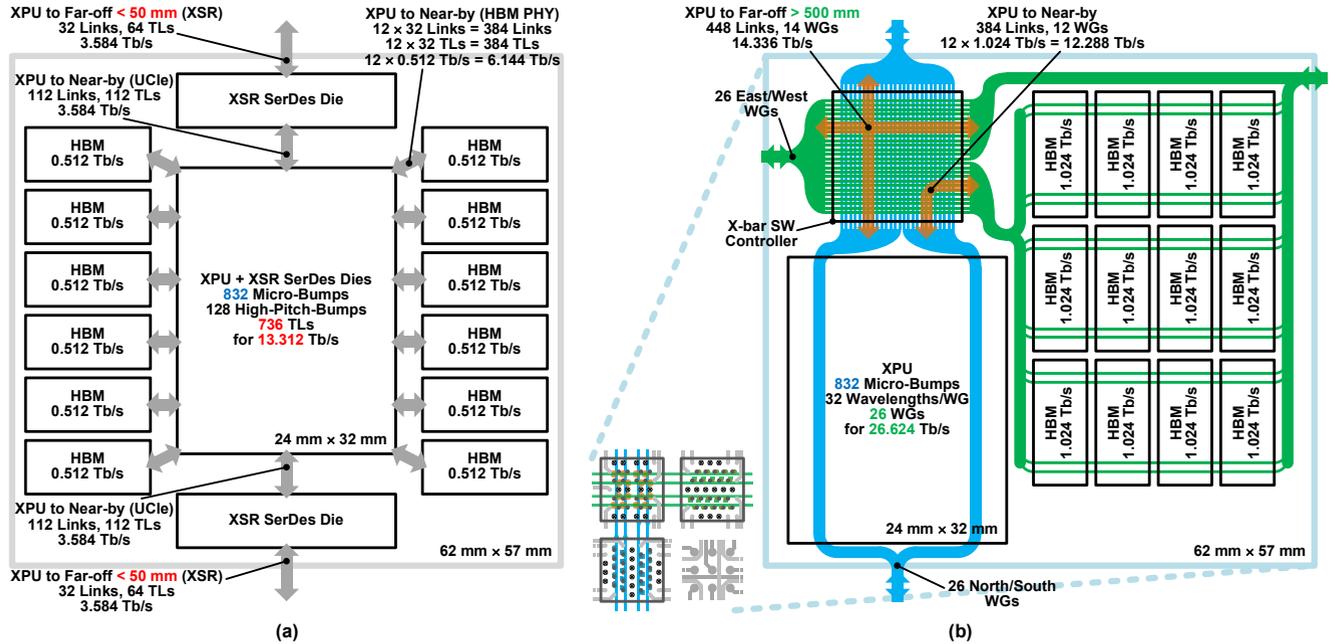

**FIGURE 3.** (a) A silicon unit-interposer containing one XPU and twelve 3D-HBM stacks while communicating through the nearby HBM PHY, nearby UCIe PHY, and far-off XSR SerDes PHY (< 50 mm) standards. (b) A photonic unit-interposer containing one XPU and twelve 3D-HBM stacks while communicating through unified optical WDM data links, regardless of nearby or far-off (> 500 mm) communication distance.

In the scenario of the proposed photonic unit-interposer shown in Fig. 3(b), the mapping between the practical scale drawing and abstract symbolic drawing, at the bottom-left corner of Fig. 3(b) used for the panel- or wafer-scale GIP top views in Figs. 1 and 2, shows that the four quadrants of the photonic unit-interposer are unequally arranged in reality, where the twelve 3D-HBM stacks occupy the first quadrant, the XPU occupies the third quadrant, and the X-bar SW controller occupies the second quadrant, on the surface of the photonic unit-interposer. The different EIC integration floor plans between the silicon and photonic unit-interposers are primarily due to their individual interconnect optimizations, based on the nature of one-to-one links on individual electrical TLs and multiple-to-multiple links on common optical WGs, respectively. In both scenarios, the EIC computing requirements per unit-interposer are identical, so their comparisons are all under the same amount of EIC die areas, unit-interposer areas, and the number of micro-bumps for data communication, as mentioned and summarized in Fig. 3. In the proposed photonic unit-interposer, the XPU communicates with all its nearby and far-off EICs through a unified optical WDM communication technique [42] with eight hundred and thirty-two 32-Gb/s single-ended electro-optical interfaces and data links [43], [44], [45], [46], [47], [48], [49] supporting photonic interconnect distances longer than 500 mm.

Overall, in the photonic unit-interposer, the total data bandwidth of the XPU is 26.624 Tb/s (= 832 links × 32 Gb/s/link) directly connected by optical WDM data links, which are all single-ended signaling I/Os with one 40-µm pitch micro-bump per link. That is, a total of 832 micro-bumps in the XPU are required for all the data links. However, each optical comb source [49], [50] can support 32 wavelengths for each WG, while one wavelength ($\lambda$) is responsible for carrying the information for one optical WDM data link; therefore, a total of 26 WGs (= 832 links ÷ 32-$\lambda$/WG) are required for all nearby and far-off data links. As shown in Fig. 3(b), the entire 26.624 Tb/s data bandwidth of the XPU is brought to the X-bar SW area through the 26 north/south WGs to perform an area-efficient and reconfigurable 0-degree or 90-degree planar redirection on a per WG basis, i.e., when the whole data information carried by the 32 wavelengths in each WG enters the X-bar SW area from any one of the four edges, the whole data information per WG can be redirected into any one of the remaining three edges and leaves the X-bar SW area.

**TABLE 1. DATA-LINK REQUIREMENTS OF SILICON AND PHOTONIC UNIT-INTERPOSERS SHOWN IN FIGS. 3(A) AND 3(B), RESPECTIVELY**

| Silicon Unit-Interposer | | Photonic Unit-Interposer | |
|---|---|---|---|
| XPU to Nearby (HBM PHY) | | XPU to Nearby (32-λ/WG WDM) | |
| # of HBM stacks | 12 | # of HBM stacks | 12 |
| # of links per HBM stack | 32 | # of links per HBM stack | 32 |
| Total # of links | 384 (= 12×32) | Total # of links | 384 (= 12×32) |
| Total # of micro-bumps | 384 (= 384×1, single-ended) | Total # of micro-bumps | 384 (= 384×1, single-ended) |
| Total # of TLs | 384 (= 384×1, single-ended) | Total # of WGs | 12 (= 384/32) |
| Data rate per link (Gb/s) | 16 | Data rate per link (Gb/s) | 32 |
| TX/RX average energy efficiency (pJ/b) | 1.01 [26] – [28] | TX/RX average energy efficiency (pJ/b) | 1.15 [43] – [47] |
| TX/RX average area per link (mm$^2$) | 0.03525 [26] – [28] | TX/RX average area per link (mm$^2$) | 0.04 [43] – [47] |
| Total bandwidth of links (Tb/s) | 6.144 (= 384×16/1000) | Total bandwidth of links (Tb/s) | 12.288 (= 384×32/1000) |
| Total power of links (W) | 6.205 (= 1.01×6.144) | Total power of links (W) | 14.131 (= 1.15×12.288) |
| Total area of links (mm$^2$) | 13.536 (= 384×0.03525) | Total area of links (mm$^2$) | 15.36 (= 384×0.04) |
| XPU to Far-off, Step 1 (UCIe PHY) | | XPU to Far-off (32-λ/WG WDM) | |
| # of XSR SerDes dies | 2 | | |
| # of links per XSR SerDes die | 112 | | |
| Total # of links | 224 (=2×112) | Total # of links | 448 |
| Total # of micro-bumps | 224 (= 224×1, single-ended) | Total # of micro-bumps | 448 (= 448×1, single-ended) |
| Total # of TLs | 224 (= 224×1, single-ended) | Total # of WGs | 14 (= 448/32) |
| Data rate per link (Gb/s) | 32 | Data rate per link (Gb/s) | 32 |
| TX+RX energy efficiency (pJ/b) | 0.6 [32], [33] | TX/RX average energy efficiency (pJ/b) | 1.15 [43] – [47] |
| TX+RX area per link (mm$^2$) | 0.0045 [32], [33] | TX/RX average area per link (mm$^2$) | 0.04 [43] – [47] |
| Total bandwidth of links (Tb/s) | 7.168 (= 224×32/1000) | Total bandwidth of links (Tb/s) | 14.336 (= 448×32/1000) |
| Total power of links (W) | 4.301 (= 0.6×7.168) | Total power of links (W) | 16.486 (= 1.15×14.336) |
| Total area of links (mm$^2$) | 1.008 (= 224×0.0045) | Total area of links (mm$^2$) | 17.92 (= 448×0.04) |
| XPU to Far-off, Step 2 (XSR SerDes PHY) | | WDM Carrier Power Injected from Optical Comb Sources | |
| # of XSR SerDes dies | 2 | Total # of WGs for data links | 26 (= 12+14) |
| # of links per XSR SerDes die | 32 | Total # of WDM carriers for data links | 832 (= 26×32) |
| Total # of links | 64 (= 2×32) | Power per WDM carrier (mW) | 0.5 (= -3.01dBm) |
| Total # of micro-bumps for UCIe | 224 (= 2×112) | Total WDM carrier power for data links (W) | 0.416 (= 832×0.5/1000) |
| Total # of high-pitch bumps for XSR | 128 (= 64×2, differential) | | |
| Total # of TLs | 128 (= 64×2, differential) | | |
| Data rate per link (Gb/s) | 112 | | |
| TX/RX average energy efficiency (pJ/b) | 1.1 [35] – [41] | | |
| TX/RX average area per link (mm$^2$) | 0.12 [35] – [41] | | |
| Total bandwidth of links (Tb/s) | 7.168 (= 64×112/1000) | | |
| Total power of links (W) | 7.885 (= 1.1×7.168) | | |
| Total area of links (mm$^2$) | 7.68 (= 64×0.12) | | |
| Summary | | Summary | |
| Total # of micro-bumps for data links | 832 (= 384+224+224) | Total # of micro-bumps for data links | 832 (= 384+448) |
| Total # of TLs for data links | 736 (= 384+224+128) | Total # of WGs for data links | 26 (= 12+14) |
| Total data bandwidth per XPU (Tb/s) | 13.312 (= 6.144+7.168) | Total data bandwidth per XPU (Tb/s) | 26.624 (= 12.288+14.336) |
| Total power of data links (W) | 18.391 (= 6.205+4.301+7.885) | Total power of data links (W) | 31.033 (= 14.131+16.486+0.416) |
| Total area of data links (mm$^2$) | 22.224 (= 13.536+1.008+7.68) | Total area of data links (mm$^2$) | 33.28 (= 15.36+17.92) |
| Energy efficiency of data links (pJ/b) | 1.38 (= 18.391/13.312) | Energy efficiency of data links (pJ/b) | 1.17 (= 31.033/26.624) |
| Bandwidth density of data links (Tb/s/mm$^2$) | 0.60 (= 13.312/22.224) | Bandwidth density of data links (Tb/s/mm$^2$) | 0.80 (= 26.624/33.28) |
| Power density of data links (W/mm$^2$) | 0.83 (= 18.391/22.224) | Power density of data links (W/mm$^2$) | 0.93 (= 31.033/33.28) |
| Maximum far-off interconnect distance (mm) | < 50 | Maximum far-off interconnect distance (mm) | > 500 |

In short, although the electrical TL routes, in terms of pitches and turning radii, in the silicon interposer have finer resolution than the photonic WG routes do, under the same amounts of XPU micro-bump usage and unit-interposer area, the high-connectivity WGs (i.e., 32 data links per WG) and the area-efficient X-bar SWs use only 26 WGs for larger than 500-mm long all-planar direction reconfigurable interconnects in the photonic interposer rather than 736 TLs for 1.5-mm to 50-mm hard-wired interconnects in the silicon interposer.

### B. BANDWIDTH & AI WORKLOAD SCALABILITIES

Continuing on the comparisons based on the unit-interposers of both scenarios shown in Fig. 3 under the identical number of micro-bump usage (i.e., 832), EIC die area (i.e., 1776 mm$^2$), and unit-interposer area (i.e., 3534 mm$^2$), as well as comparable average energy efficiency of data links (i.e., 1.38 pJ/b vs. 1.17 pJ/b elaborated in the next subsection), the overall data bandwidth per XPU (i.e., 13.312 Tb/s) in the silicon unit-interposer has been confined by the number of electrical TLs (i.e., 736) that can be accommodated within the physical space of the silicon interposer under the constraints of signal integrity, channel loss, crosstalk, power delivery, and transceiver power density associated with heat dissipation requirements. On the other hand, the overall data bandwidth per XPU (i.e., 26.624 Tb/s) in the photonic unit-interposer still has potential to be further enhanced with the development of silicon-photonics process technology, including the parasitic capacitance and quality factor (Q) of micro-ring resonators (MRR) [51], thermal control efficiency [52], [53], [54], PD responsivity [55], and linearities for both E/O and O/E conversions [56], [57]. Therefore, the data rate per optical data link (i.e., 32 Gb/s) can be further scaled up without compromising the energy efficiency of data links and the required numbers of WGs and WDM wavelengths.

Regarding the scalability of AI workloads, although multiple silicon unit-interposers can be tiled together through direct semiconductor manufacturing on a single silicon wafer, the scalability of an AI computing system integrated on this wafer-scale silicon interposer is limited by two main factors. One is the achievable wafer size (i.e., a 30-cm or 12-inch diameter) in today's semiconductor process technology, which has shown minimal progress over the past two decades [24], as mentioned. The other is that networking diversity is limited to adjacent unit-interposers, meaning that within a wafer-scale silicon interposer, there is no direct far-off chiplet-to-chiplet communication beyond adjacent EICs, and computing resource utilization can't be dynamically optimized. On the other hand, the panel-scale PIC-embedded GIP can tile up the photonic unit-interposers unlimitedly with the workload requirement of an AI computing system, while the wafer-scale MPH-integrated GIP can tile the photonic unit-interposers up to the size of a single silicon wafer. In any case of the photonic interposer, the reconfigurability described in the previous subsection enables broadband data communication between any two EICs within the entire panel-scale or wafer-scale AI computing system, thereby maximizing networking diversity and dynamic utilization of computing resources.

### C. ENERGY EFFICIENCY & BANDWIDTH DENSITY

In the scenario of the silicon unit-interposer, the energy efficiency, bandwidth density, and power density of the data links are the average results contributed by all three I/O standards, i.e., HBM, UCIe, and XSR SerDes PHYs, in terms of the total data bandwidth per XPU (i.e., 13.312 Tb/s), the total power consumption of the data links (i.e., 18.391 W), and the total on-chip area of the data links (i.e., 22.224 mm$^2$), whose detailed estimation process, numbers, and reference literature are summarized on the left side of Table 1. On the other hand, in the scenario of the photonic unit-interposer, the energy efficiency, bandwidth density, and power density of the data links are the average results contributed by the unified optical WDM data links and the required WDM carrier power injected from either external optical comb sources or in-package integrated optical comb sources, as shown in Fig. 1 or 2 and discussed in Section IV. According to the detailed estimation process, numbers, and reference literature summarized on the right side of Table 1, the total data bandwidth per XPU is 26.624 Tb/s, the total power consumption of the data links is 31.033 W, and the total on-chip area of the data links is 33.28 mm$^2$.

Overall, the silicon and photonic unit-interposers have the identical number of micro-bump usage (i.e., 832), identical EIC die area (i.e., 1776 mm$^2$), identical unit-interposer area (i.e., 3534 mm$^2$), comparable energy efficiency of data links (i.e., 1.38 pJ/b vs. 1.17 pJ/b), comparable bandwidth density (i.e., 0.6 Tb/s/mm$^2$ vs. 0.8 Tb/s/mm$^2$), and comparable power density (i.e., 0.83 W/mm$^2$ vs. 0.93 W/mm$^2$). Although the photonic unit-interposer has slightly better performance in all aspects except for power density, the deltas are all within the ranges of the estimation errors due to various CMOS semiconductor process nodes for the electronic parts of the data links.

In short, the analysis result indicates that when the six different specifications of the silicon and photonic unit-interposers are all comparable, the photonic unit-interposer outperforms the silicon unit-interposer in terms of twofold data bandwidth per XPU (i.e., 13.312 Tb/s vs. 26.624 Tb/s), with much fewer physical interconnect routes (i.e., 736 TLs vs. 26 WGs) and much longer interconnect distances (i.e., < 50 mm vs. > 500 mm).

## IV. BUILDING BLOCKS OF GLASS-SUBSTRATE PHOTONIC INTERPOSERS

The key building blocks of the PIC-embedded and MPH-integrated GIPs are elaborated in this section, in terms of their design theories, functionalities, simulations, and/or preliminary experimental validations.

### A. OPTICAL TRANSCEIVERS

The E/O interface of WDM data links per WG is shown in Fig. 4(a), where the EIC-to-interposer physical connections are at the touchpoints between the EIC micro-bumps and PIC micro-pads of the PIC-embedded GIP in Fig. 1 or GIP micro-pads of the MPH-integrated GIP in Fig. 2, respectively. Each optical TX in the EIC is implemented by high-speed serializers with AC-coupled high-dynamic-range 4-level pulse-amplitude-modulation (PAM-4) drivers running at higher than 32-Gb/s data rates [43], [44], [45], [48], [49], [58]. The series inductor ($L_S$) at the TX output is used to resonate with the parasitic capacitance at each micro-bump location, mainly contributed by the electrostatic discharge diodes ($C_{ESD}$), for the bandwidth extension of each E/O interface. The DC conductivity from the TX is intentionally blocked at the micro-pad of the PIC or GIP. Still, the broadband AC conductivity from the TX is continued through the AC-coupling capacitance ($C_{AC}$) formed by the micro-pad and top-metal layers in the PIC or GIP with a high dielectric-constant silicon-nitride (SiN) layer in between [9]. The vertical metal/via stack below the top-metal layers conducts the broadband AC signal (i.e., data information) and merges with a high-resistivity DC signal path ($R_{DC}$) in the PIC or GIP before reaching the central electrode of the proposed silicon-rich-nitride (SRN) MRR, described in Section IV-C, whose effective refractive index ($n_{eff}$) variation with the broadband AC signal amplitude can be enhanced with the strength of the DC electric field between the central and peripheral electrodes of the SRN MRR for improving the optical modulation amplitude (OMA), as well as extinction ratio (ER), for the E/O conversion. Overall, this 3D heterogeneous integration structure from the TX output, serial-peaking inductor, micro-bump, micro-pad, broadband AC-coupling capacitor, and metal/via stack to the SRN MRR central electrode forms the minimal EIC-to-interposer interconnect distance as today's advanced heterogeneous integration technology [44], [45], [58], [59] with the vertical alignments of all signal-conductive components on the same planar coordinate, which can not only minimize the high-frequency electrical loss but also form a lumped signal transmission path for the DC high-impedance terminated SRN MRR electrode with almost zero DC power consumption. Note that the DC voltages for the SRN MRR central and peripheral electrodes, as well as the EIC power supplies, biases, and clocks from the external sources, are provided through the metal/via stacks, TSVs, and bumps of the PIC or GIP, as shown in Figs. 1(a), 2(a), and 4(a), as well as the RDL assembled underneath the PIC or GIP.

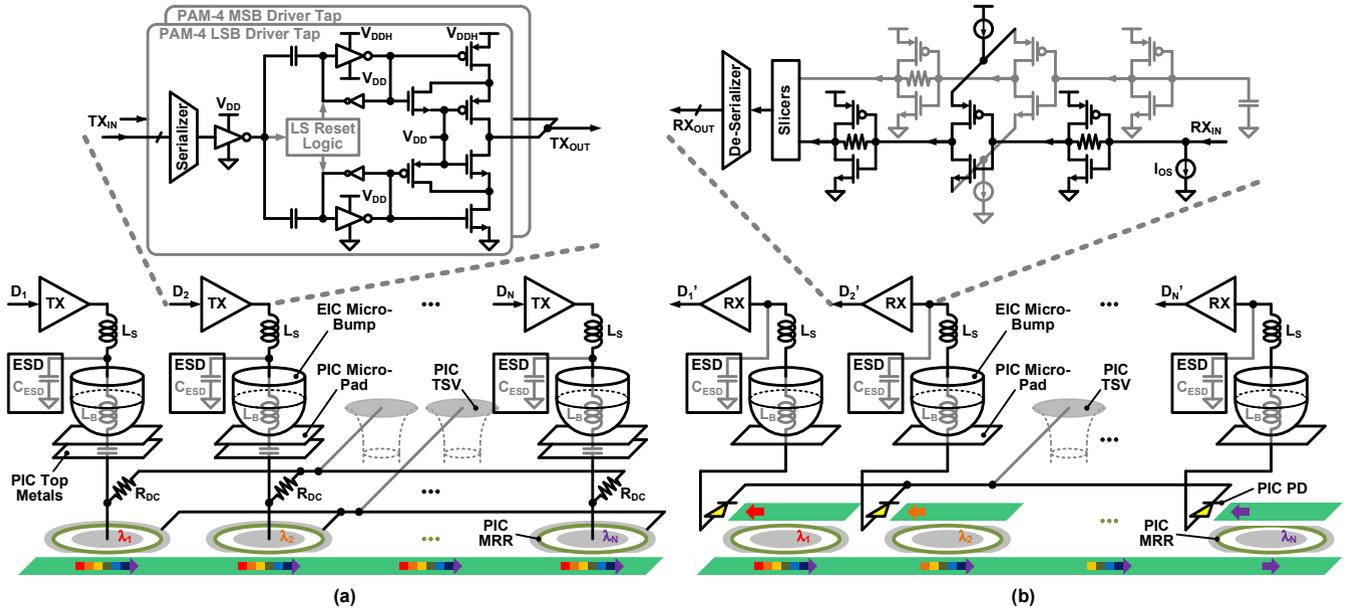

**FIGURE 4.** (a) The WDM E/O interface, including optical TXs, EIC micro-bumps, PIC (or GIP) micro-pads, PIC (or GIP) SRN MRRs for intensity modulations, and DC biasing TSVs. (b) The WDM O/E interface, including optical RXs, EIC micro-bumps, PIC (or GIP) micro-pads, PIC (or GIP) SRN MRRs for WDM channel selections, PIC (or GIP) PDs, and DC biasing TSVs.

The O/E interface of WDM data links per WG is shown in Fig. 4(b), where the interposer-to-EIC physical connections are at the touchpoints between the PIC or GIP micro-pads and EIC micro-bumps. Similar to the E/O interface but in an opposite signal propagation direction, the 3D heterogeneous integration structure from the SRN MRR drop-port output, PD, metal/via stack, micro-pad, micro-bump, serial-peaking inductor, to the optical RX input also forms the minimal interposer-to-EIC interconnect distance as today's advanced heterogeneous integration technology [46], [47], [59] with the vertical alignments of all signal-conductive components on the same planar coordinate, which can minimize the high-frequency electrical loss and meanwhile form a lumped signal transmission path without the concern of impedance matching. The power supply for the PDs from the external source is also provided through the metal/via stacks, TSVs, and bumps of the PIC or GIP, as shown in Figs. 1(a), 2(a), and 4(b), as well as the RDL assembled underneath the PIC or GIP.

Each optical RX in the EIC is implemented by a linear analog front-end (AFE) to convert the received serial data from the photocurrent into the voltage domain, followed by strong-arm-latch-based PAM-4 slicers associated with data de-serialization functionality. This AFE consists of three sub-amplifiers in series [46], [47], [60], [61]: a trans-impedance amplifier (TIA), a trans-admittance amplifier (TAA), and another trans-impedance amplifier (TIA). Each TIA is a voltage-to-current feedback architecture whose feedforward path is formed by complementary P-type and N-type transconductance stages, and both share a single DC bias current path from the analog supply to the ground. The feedback path of each TIA is formed by a tunable resistor, which plays a key role in setting the TIA gain, bandwidth, and input/output impedance across all different process, voltage, and temperature (PVT) conditions [60], [61]. The differential TAA and its push-pull tail currents render the single-ended to differential functionality while a pair of TIAs follow the TAA differential output; cascading the TAA and TIA forms a broadband high-swing linear voltage-to-voltage amplifier (or a Cherry-Hooper amplifier [62]) since the TAA and TIA perform a voltage-to-current and a current-to-voltage conversion, respectively. This TAA + TIA architecture possesses two major advantages: first, the separation of the TAA and TIA stages greatly improves the AFE output dynamic range and linearity; second, the TIA not only increases the bandwidth of the TAA due to its low AC input impedance but also provides DC common-mode feedback, allowing the complementary transconductance in the TAA without additional power consumption.

### B. OPTICAL CROSSBAR SWITCHES
A conventional add-drop micro-ring resonator (AD-MRR) is formed by a single-layer waveguide structure, as shown on the left of Fig. 5(a), with transverse electric (TE) mode coupling between the add/drop waveguides and ring waveguide to perform light propagation from the add-port to the drop-port.

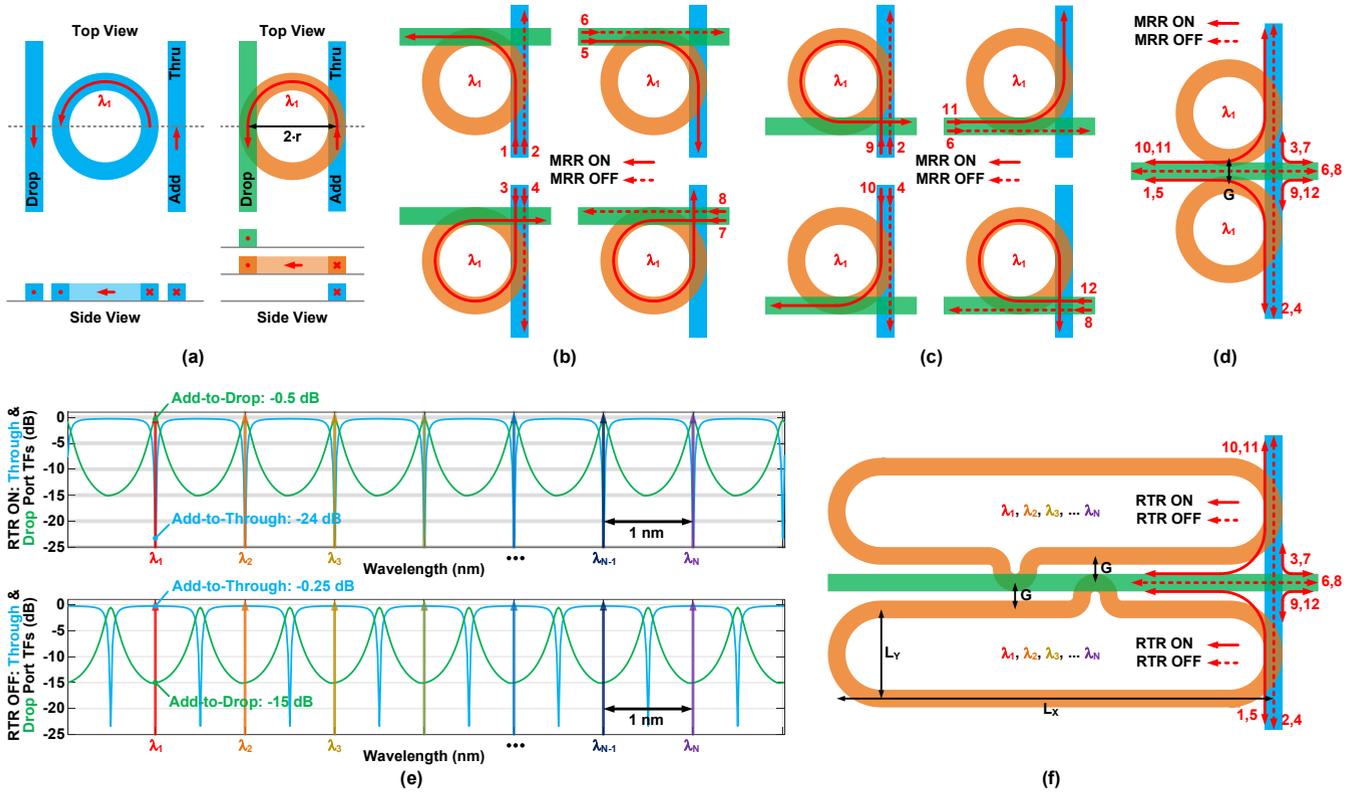

**FIGURE 5.** (a) A conventional AD-MRR (left) and the proposed VC-AD-MRR (right). (b) and (c) show two VC-AD-MRR positions to cover all 12 in-and-out propagations per crossing node. (d) A complete X-bar SW formed by two VC-AD-MRRs per crossing node for the planar redirection of a single WDM data link per WG. (e) The add-to-drop and add-to-through TFs of a VC-AD-RTR in the RTR-ON or RTR-OFF modes. (f) A complete X-bar SW formed by two VC-AD-RTRs per crossing node for the simultaneous planar redirection of multiple WDM data links per WG.

Under a similar concept, this paper exploits multi-layer SiN waveguides, high-precision $SiO_2$ planarization, and high-tolerance waveguide-alignment manufacturing capability [9] with TE-mode coupling to perform a vertically-coupled add-drop MRR (VC-AD-MRR) to enable low-loss vertical (i.e., perpendicular to the substrate) bi-directional up-down light-signal transition between the multi-layer waveguides as shown on the right of Fig. 5(a). With this technique, VC-AD-MRRs can be responsible for switching 90-degree planar directions of light, as shown in Figs. 5(b) and 5(c), at each crossing node of the WGs shown in Figs. 1(b), 2(b), and 3.

The directional switching capability is controlled through the static thermal control of each VC-AD-MRR; when the thermal control voltage sets the critical coupling wavelength of the VC-AD-MRR to the light-signal carrier wavelength (e.g., $\lambda_1$), the direction of the incoming light signal from the add-port is turned by 90 degrees (i.e., MRR ON) to the drop-port; otherwise, it is passed through the crossing node from the add-port to the through-port (i.e., MRR OFF). Ideally, an X-bar SW should support all twelve in-and-out propagation paths, i.e., incoming from any one of the north, south, east, or west directions and outgoing from any one of the remaining three directions. However, a single VC-AD-MRR, for example, sitting at the southwestern corner of a crossing node as shown in Fig. 5(b), can only support eight in-and-out propagation paths (i.e., path # 1, 2, 3, 4, 5, 6, 7, and 8) due to the nature of light coupling among the waveguides (from blue, orange, to green, and vice versa). Therefore, another identical VC-AD-MRR sitting at the northwestern corner of a crossing node, as shown in Fig. 5(c), is necessary to support the other set of eight in-and-out propagation paths (i.e., path # 2, 4, 6, 8, 9, 10, 11, and 12) so that two VC-AD-MRRs together, serving as a single X-bar SW per crossing node, as shown in Fig. 5(d), can fully cover all twelve in-and-out propagation paths.

To support multi-wavelength X-bar switching capability as mentioned in Sections II and III, the VC-AD-MRR technique can be further expanded into a vertically-coupled add-drop race-track resonator (VC-AD-RTR) to simultaneously switch the broadband WDM optical signals carried by multiple wavelengths ($\lambda_1, \lambda_2, …, \lambda_N$) at each crossing node, as shown in Fig. 5(f). To simultaneously affect the propagation directions of all 32 data links carried by 32 individual wavelengths in a single WG, the resonance cavity of the RTR has to be at least 32-fold longer than the MRRs used for the E/O and O/E conversions described in Section IV-A to correspondingly generate a 1/32-fold smaller free spectral range (FSR) of resonance wavelength spacing for matching the WDM wavelength spacing (e.g., about 1 nm), as shown in the spectra of Fig. 5(e) [48], [49], [63], across all 32 carriers, which requires the replica design and thermal control offset trimming [64] between the VC-AD-RTRs in the X-bar SW and the RTR used for the optical comb generation described in Section IV-D. Also, it is important to note that although the VC-AD-RTRs in Fig. 5(f) requires a larger X-bar SW area than the VC-AD-MRRs in Fig. 5(d), the VC-AD-RTR not only offers broadband switching rather than a single data link but also resolves the unwanted mutual resonance between the two MRRs per X-bar SW shown in Fig. 5(d), where the distance "G" between the MRRs needs to be large enough to avoid the mutual resonance between each other while weakens the add-to-drop coupling between each MRR and the common WG colored in green. On the other hand, the longer route of the RTRs shown in Fig. 5(f) allows for routing flexibility, keeping the distance "G" between the two RTRs without sacrificing the coupling quality between each RTR and the green WG.

The simulated add-to-through and add-to-drop power transfer functions (TF) of a VC-AD-RTR, based on multi-layer SiN WG geometry in a PIC process design kit (PDK) [9], are shown in Fig. 5(e) under two different thermal control conditions to enable either add-to-drop (RTR ON) or add-to-through (RTR OFF) coupling for all data links carried by their individual WDM carriers ($\lambda_1$, $\lambda_2$, ..., $\lambda_N$). As mentioned, the FSR of the VC-AD-RTR matches the FSR of the RTR in the optical comb source generating the WDM carriers; therefore, in the RTR-ON mode, all the WDM carriers can propagate from a blue WG, for example, to a green WG through the low-loss (-0.5 dB) add-to-drop TF, and equivalently perform a 90-degree turn at the crossing node between the blue and green WGs, while the high-loss (-24 dB) add-to-through TF blocks all the WDM carriers staying on the blue WG. In contrast, in the RTR-OFF mode, all the WDM carriers can propagate and stay in a blue WG, for example, through the low-loss (-0.25 dB) add-to-through TF and equivalently perform a 0-degree bypass at the crossing node without leaking to the green WG because of the high-loss (-15 dB) add-to-drop TF.

Finally, according to the losses and numbers of the X-bar SWs, as discussed and shown in Fig. 5(e) in either RTR-ON or RTR-OFF mode, within a photonic interconnect between two EICs, the required injection power per WDM carrier from the optical comb source can be calculated. For example, the brown signal path between EIC-1 and EIC-2 in Fig. 1(b) contains one X-bar SW performing a 0-degree bypass (i.e., add-to-through TF) and two X-bar SWs performing a 90-degree turn (i.e., add-to-drop TF), so the brown signal path has at least a 1.25-dB (= 1×0.25-dB + 2×0.5-dB) loss due to the X-bar SWs.

Based on the X-bar SW reconfigurability and scalability of the panel-scale PIC-embedded GIP discussed in Section III, assuming an in-package AI computing system is composed of 64 tiles (i.e., 8-by-8) of the photonic unit-interposer shown in Fig. 3(b), then one of the worst-case scenarios (i.e., the highest photonic interconnect loss) is that an EIC located at the bottom-left corner tile is communicating with another EIC located at the top-right corner tile, whose photonic interconnect contains twelve 0-degree-bypass and two 90-degree-turn X-bar SWs. To satisfy -10-dBm optical RX sensitivity with a 0.8-A/W PD responsivity in general [46], [47] and additional losses due to the edge couplers and WGs, the required carrier power injected from the optical comb source in this example should be around -3 dBm (= -10-dBm + 12×0.25-dB + 2×0.5-dB + 3-dB, where the extra 3 dB is counted for the losses of edge couplers and WGs). That is about 0.5 mW per WDM carrier, which has been included in the power consumption of the photonic unit-interposer listed in Table 1.

### C. SILICON-RICH-NITRIDE MICRO-RING RESONATORS ON GLASS SUBSTRATE

The state-of-the-art lithium niobate ($V_\pi L \approx 1.8$ V·cm) and other high-k dielectrics like barium titanate offer high optical intensity modulation efficiencies for the E/O conversion, but they face significant challenges: lack of CMOS compatibility, low refractive indices requiring larger waveguides, and high RF permittivity reducing the modulation electric fields [65], [66], [67], [68], [69], [70]. Consequently, silicon-based carrier dispersion modulators ($V_\pi L \approx 3.5$ V·cm) remain prevalent despite high dopant-related propagation losses [71]. However, these modulators rely on processing silicon substrates, which may not be further enlarged up to panel scales for integrating an extremely high-workload AI computing system. Also, a CMOS-compatible alternative remains elusive due to the limited nonlinear optical properties of most materials, including stoichiometric SiN. Therefore, the development of an active nonlinear material platform that can meet the need for substrate scalability remains a challenge.

Recent research has shown that SRN films can exhibit enhanced nonlinear susceptibility coefficients, $\chi^{(2)}$ and $\chi^{(3)}$, with increased silicon content, even in low-temperature plasma-enhanced chemical vapor deposition (PECVD) processes [72]. In this paper, a $\chi^{(3)}$-based linearized electro-optic SRN MRR using heterodyne gain is proposed for high-speed E/O conversion on a glass substrate and demonstrated by systematic performance evaluation, numerical studies, and experimental validations.

For SRN-based modulators, theoretically, the $V_\pi L$ metrics are about 2 to 3.5 V·cm, and the V$\pi$L·$\alpha$ metrics are about 116 V·dB in phase modulators. With further optimization for intensity modulation in Mach-Zehnder interferometers (MZI) ($V_\pi L < 1$ V·cm) and MRRs, the results show that a linearized $\chi^{(3)}$-based modulator is a viable and manufacturable alternative compared to $\chi^{(2)}$-limited and plasma-dispersion silicon modulators. Thus, the third-order non-resonant electronic nonlinearities of SRN, which can enable ultrafast responses, are promising for high-speed intensity modulation and wave-mixing applications [73]. The derived expressions for refractive index changes based on $\chi^{(2)}$, $\chi^{(3)}$, and a combination of $E_{dc}$ and $E_{ac}$ fields are summarized in Table 2, which highlights the key properties of an arbitrary material under static ($E_{dc}$) and time-varying ($E_{ac}$) electric fields. The static refractive index change is given by

$$\Delta n_{dc} = \Delta n_{dc}^{\chi^{(2)}} + \Delta n_{dc}^{\chi^{(3)}} \qquad (1)$$

while the modulated component is

$$\Delta n_{ac} = \Delta n_{ac}^{\chi^{(2)}} + \Delta n_{ac}^{\chi^{(3)}} + \Delta n_{ac+dc}^{\chi^{(3)}} \qquad (2)$$

For $\chi^{(3)}$-based linearized modulators, $\Delta n_{ac}$ is critical with the mixed term, i.e., the third term in Eq. (2), offering two key advantages: first, when $E_{dc} \gg E_{ac}$, it is approximately linear in $E_{ac}$, which can reduce quadratic chirping; second, it provides a form of heterodyne gain, which can allowing a weaker $E_{ac}$ field to achieve the desired modulation when paired with a strong $E_{dc}$ field. These are beneficial for the proposed applications, where high DC voltages are feasible, but low AC voltages are often required. With these, the estimation of effective $\chi^{(2)} = 3$ and $\chi^{(3)} E_{dc} = 180$ pm/V for SRN at $E_{dc} = 10^8$ V/m, which surpasses the 27 pm/V of Lithium Niobate. Meanwhile, SRN also features a high refractive index up to 3.1 with measured enhanced $\chi^{(3)}$ up to $12.6 \times 10^{-19}$ m$^2$/V$^2$ [74], high breakdown fields, and low optical losses over broad spectral ranges. These attributes make SRN an attractive candidate for linearized modulators based on the DC-induced Pockels effect.

**TABLE 2. THE SECOND AND THIRD ORDER NONLINEAR CONTRIBUTIONS TO THE CHANGE OF REFRACTIVE INDEX [22]**

| | $\chi^{(2)}$ | $\chi^{(3)}$ |
|---|---|---|
| DC | $\Delta n_{dc}^{\chi^{(2)}} = \sum_{jk} \frac{\chi_{ijk}^{(2)}}{n_k^{eq}} E_j^{dc}$ | $\Delta n_{dc}^{\chi^{(3)}} = \sum_{jk} \frac{3\chi_{ijjk}^{(3)}}{2n_k^{eq}} E_j^{dc\,2}$ |
| AC | $\Delta n_{ac}^{\chi^{(2)}} = \sum_{jk} \frac{\chi_{ijk}^{(2)}}{n_k^{eq}} E_j^{ac}$ | $\Delta n_{ac}^{\chi^{(3)}} = \sum_{jk} \frac{3\chi_{ijjk}^{(3)}}{2n_k^{eq}} E_j^{ac\,2}$ |
| AC + DC | Not Applicable | $\Delta n_{ac+dc}^{\chi^{(3)}} = \sum_{jk} \frac{3\chi_{ijjk}^{(3)}}{n_k^{eq}} E_j^{ac} E_j^{dc}$ |

Fig. 6 presents the simulation results of the DC electric field distribution across the SRN waveguide geometry. Key design parameters, such as the electrode gap ($d_{gap}$) and waveguide gap ($w_{gap}$), can be adjusted to optimize the device performance. Reducing these parameters enhances the strength of the DC electric field but simultaneously increases optical losses. Another critical factor is the choice of the cladding material. In this simulation, SiO$_2$ was used due to its widespread availability and compatibility; however, it has a relatively low dielectric constant compared to high-$k$ dielectrics such as Al$_2$O$_3$ or HfO$_2$, which could further enhance the electric field. Overall, the average electric field ($E_x$) in the SRN waveguide is 1.5 V/μm when a 5-V DC is applied under $w_{gap} = 600$ nm, $d_{gap} = 300$ nm. This implies that a voltage of approximately 330 V DC needs to be applied to reach the voltage breakdown field of SRN, which is approximately 100 V/μm, as determined in prior work [75]. In that case, the effective $\chi^{(2)} = 3$ and $\chi^{(3)} E_{dc} = 180$ pm/V, which is much larger than that of LiNbO$_3$ as mentioned. Importantly, it has been demonstrated that SiN can be poled [76], and therefore, there is no need to maintain a large DC voltage during modulator operation.

One crucial advantage of SRN is its ability to permanently change its refractive index through annealing, which enables optical trimming for the proposed SRN MRR and can be performed locally using continuous-wave (CW) visible lasers, as demonstrated in [77]. Fig. 7 illustrates experimental results showing how trimming enables the resonant wavelength of each SRN MRR to be set to match one of the WDM carriers for both E/O and O/E conversions, as mentioned in Section IV-A, with an accuracy of 10 pm [77]. This property is also highly beneficial for trimming the RTRs in the X-bar SWs to align the FSR spacing with all WDM carriers per WG, as mentioned in Section IV-B. Note that the free carrier absorption (FCA) due to Si-H bonds in SRN WGs made of films with an index of 3 can lead to nonlinear optical losses [78], but this effect has not been observed for SRN films with an index of 2.5 and lower. Therefore, to minimize the loss associated with Si-H absorption, the precursor gas SiH4 can be substituted with deuterated silane SiD4 to reduce hydrogen absorption in the optical bandwidth used for the WDM data links.

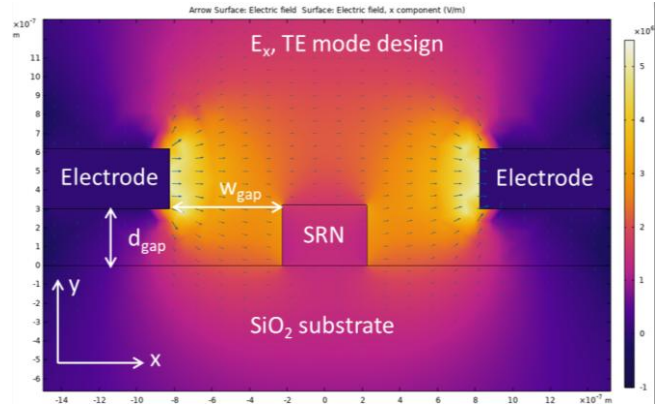

**FIGURE 6.** The DC electric field simulation of an SRN MRR, where the color distribution shows $E_x$ in V/m when 5 V is applied.

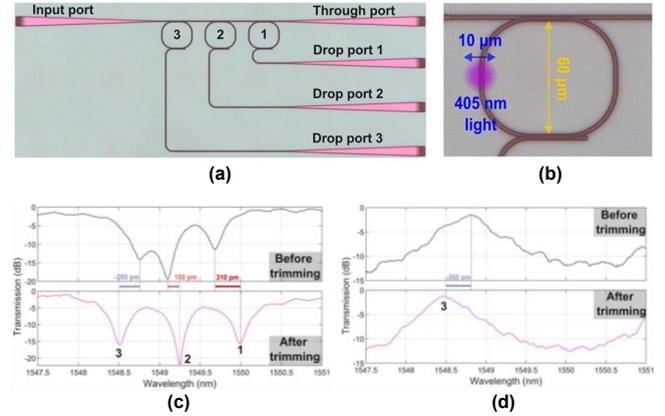

**FIGURE 7.** (a) The microscope image of three SRN MRRs. (b) The magnified microscope image of an SRN MRR. (c) The measured add-to-through TFs before and after the trimming. (d) One of the measured add-to-drop TFs before and after the trimming.

### D. SILICON-RICH-NITRIDE COMB SOURCE ON GLASS SUBSTRATE

Leveraging silicon for optical frequency comb generation is challenging due to its power handling capability, limited by two-photon absorption and free-carrier absorption. Because of the wide bandgap and ultralow loss performance, SiN is a viable alternative for manufacturing comb generators on glass substrates. The nearly 5 eV bandgap of SiN circumvents two-photon absorption at the WDM wavelengths, resulting in a transparency wavelength range of 400 nm to 2350 nm. The material has a refractive index of about 2.0 at 1550-nm wavelengths and provides a 24% index contrast relative to silicon dioxide cladding for waveguiding. SiN, however, has a nonlinear optics coefficient that is 100 times smaller than that of silicon. Its platform is weakly nonlinear and requires a millimeter on-chip footprint to implement nonlinear photonic circuits on a large glass substrate. In addition, the ultralow-loss performance of SiN requires high-temperature annealing at 1200 °C. The annealing temperature is incompatible with the thermal budget in backend-of-line CMOS manufacturing (400 °C). To overcome the challenges above, the PECVD is proposed for depositing SRN to generate optical frequency combs. By engineering the stoichiometry of SiN, SRN offers an enhancement in nonlinearity while preserving a bandgap of at least 1.7 eV, thereby circumventing two-photon absorption at 1550-nm wavelengths.

To demonstrate that SRN is a highly nonlinear optical material for optical comb generators, the preliminary measurement results of nonlinear optical coefficients for SRN at different silicon-to-nitrogen compositions are shown in this subsection. The thin film deposition is performed in a PECVD reactor with an ammonia-free chemistry at 350 °C and 650 mTorr. The silane-to-nitrogen gas flow rate ratio is tuned from 0.1 to 2.5 to control the stoichiometry. Fig. 8 demonstrates how the gas flow rate ratio changes the refractive index, bandgap, and nonlinear refractive index of SRN. In each deposition, both bare silicon substrates and fused silica substrates are loaded into the reactor. The silicon substrate samples are used to characterize the optical constants of the deposited films with variable-angle spectroscopic ellipsometry. Then, the refractive index and bandgap are extracted from fitting the ellipsometry data to the Tauc-Lorentz model, spanning wavelengths from 190 nm to 1690 nm. The z-scan method is further used to characterize the nonlinear refractive index of the SRN film deposited on the fused silica substrate, which is employed to minimize the measured nonlinearity. Additionally, the repetition rate of the femtosecond pulsed laser is set at 5 kHz to mitigate thermal lensing in z-scan measurements. The measurement results in Fig. 8 demonstrate that SRN outperforms other platforms in terms of nonlinear refractive index and bandgap.

The modeling of comb formation in an SRN RTR is performed. The WG of a circled SRN RTR with a cross section of 550 nm by 313 nm is engineered to achieve anomalous dispersion over a bandwidth exceeding 100 nm. The radius of the circled SRN RTR is chosen at 133 µm to achieve a nearly 100-GHz FSR (i.e., 0.8-nm spacing) for the WDM carrier generation. Fig. 9 demonstrates the dynamical formation of comb lines as the excitation (pump) wavelength is tuned across the cavity resonance. This wavelength detuning is a technique for accessing various operational regimes of the nonlinear resonator. The first regime at the 790th detuning step corresponds to modulational instability, providing the parametric gain in an optical resonator. In this regime, the pump energy is spontaneously converted into primary comb lines through cascaded four-wave mixing. As wavelength conversion between comb lines grows, a breathing pulse is formed. The breather acts as a stimulus to excite a stable pulse, as shown at the 1600th detuning step. The stable pulse is known as a dissipative Kerr soliton, and its formation results in the phase locking of comb lines. Therefore, Fig. 9 theoretically demonstrates the generation of phase-locked combs in the SRN RTR, given that the RTR exhibits anomalous dispersion, a loss of 1 to 2 dB/cm, and a nonlinearity of 935 $W^{-1}m^{-1}$ under an input of 300 mW.

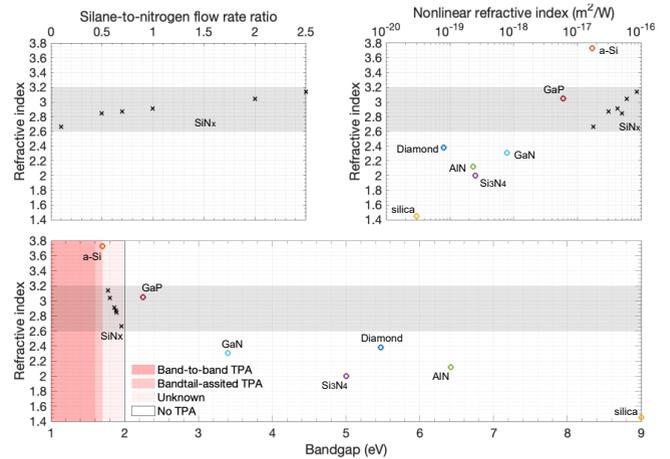

**FIGURE 8.** Thin film characterization for the nonlinear refractive index and Tauc bandgap of SRN at selected silicon-to-nitrogen compositions.

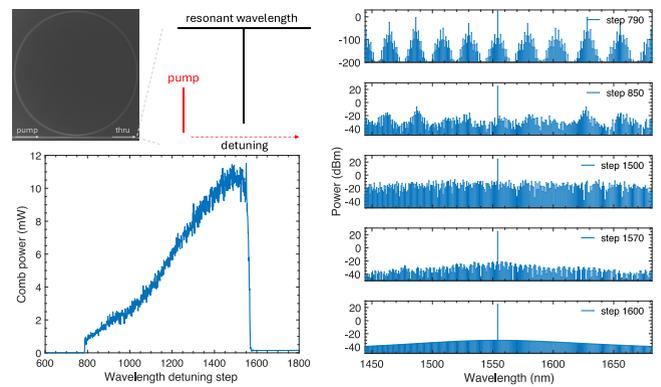

**FIGURE 9.** The modeling of comb formation in an SRN RTR.

## V. SUMMARY & FUTURE WORK

As demonstrated in this paper, photonic interposers offer high scalability, high bandwidth, reconfigurable interconnections, low energy consumption, and compelling bandwidth density for in-package AI computing system integration. The next significant advantage of in-package photonic interconnects over electronic interconnects, which has not yet been fully revealed in this paper with concrete data points, is the ability for two chiplets located at opposite ends of a panel-scale interposer to directly communicate with each other entirely in the photonic domain without intermediate O/E/O conversions or electronic repeaters. Therefore, our future work will focus on further developing several aspects as follows.

First, in today's wafer-scale silicon interposer, the communication between two chiplets on the opposite ends of the interposer would have to be forwarded through potentially many intermediate electronic chiplets, which could consume a substantial fraction of the micro-bumps and TLs available to each of the intermediate chiplets along the communication path. Additionally, a significant portion of the die area would have to be devoted to carefully engineered data repeaters and clock synchronization circuits across the chiplets in the communication path, as well as potentially the die area dedicated to switching functions if the communication path can be reconfigurable, such as the feature presented in this paper. All these factors could substantially increase the power requirements and significantly limit the achievable data bandwidth. To fully verify these limitations in today's silicon-interposer approach and showcase the full benefits of our photonic-interposer approach, detailed system-level architecture studies with in-depth application workloads would be required.

Second, as demonstrated in Section IV-B, the proposed VC-AD-MRR crossbar switch with micro-ring resonators can, in principle, be used to support per-wavelength switching, albeit with the downside of its internal mutual resonance. Therefore, this paper focuses on the alternative VC-AD-RTR crossbar switch with racetrack resonators, which supports broadband per-waveguide switching. Our future work is planned to maximize the reconfigurability feature of the proposed optical crossbar switches, which can possess both per-wavelength and per-waveguide switching capabilities without mutual resonance loss, thereby enabling a much finer photonic-interconnect granularity for future extremely high-workload AI applications.

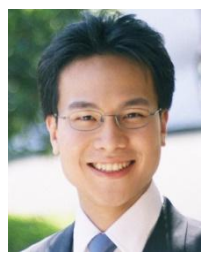

**TZU-CHIEN HSUEH** (Senior Member, IEEE) received the Ph.D. degree in Electrical and Computer Engineering from the University of California, Los Angeles (UCLA), in 2010. From 2001 to 2006, he was a Mixed-Signal Circuit Design Engineer in Hsinchu, Taiwan. From 2010 to 2018, he was a Research Scientist at Intel Lab Signaling Research and an Analog Engineer at Intel I/O Circuit Technology in Hillsboro, Oregon. Since 2018, he has been an Assistant Professor and is currently an Associate Professor in the Department of Electrical and Computer Engineering at the University of California, San Diego (UCSD). His research interests include wireline electrical/optical transceivers, clock-and-data recovery, data-conversion circuits, on-chip performance measurements/analyzers, and digital/mixed signal processing techniques.

Dr. Hsueh was a recipient of multiple Intel Division and Academy Awards from 2012 to 2018, the 2015 IEEE Journal of Solid-State Circuits (JSSC) Best Paper Award, the 2020 NSF CAREER Award, and the 2022 UCSD Best Teacher Award. From 2016 to 2018, he served on the Patent Committee for Intel Intellectual Property (Intel IP) and the Technical Committee for the Intel Design & Test Technology Conference (DTTC). From 2018 to 2024, he has served on the Technical Program Committee for the IEEE Custom Integrated Circuits Conference (CICC) and as a Guest Associate Editor for IEEE Solid-State Circuits Letters (SSC-L).

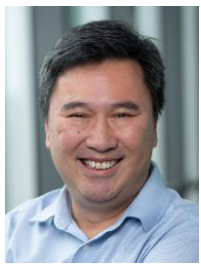

**BILL LIN** (Senior Member, IEEE) received the B.S., M.S., and Ph.D. degrees in electrical engineering and computer sciences from the University of California at Berkeley, Berkeley, CA, USA, in 1985, 1988, and 1991, respectively. He is currently a Professor in electrical and computer engineering with the University of California, San Diego, La Jolla, CA, USA, where he is actively involved with the Center for Wireless Communications (CWC), the Center for Networked Systems (CNS), and the Qualcomm Institute in industry-sponsored research efforts. Dr. Lin's research has led to more than 200 journal and conference publications, including a number of Best Paper awards and nominations. He also holds five awarded patents. He was the General Chair and on the executive and technical program committee of many IEEE and ACM conferences, and he was Associate and Guest Editors for several IEEE and ACM journals.

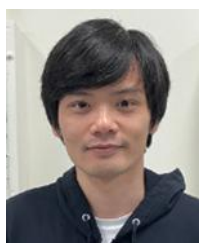

**ZIJUN CHEN** (Graduate Student Member, IEEE) received the B.S. degree in Electrical Engineering from the University of California, San Diego. He is currently pursuing Ph.D. in Electrical Engineering (Photonics) with the Ultrafast and Nanoscale Optics group at the University of California, San Diego. His research explores integrated nonlinear optics for generation, manipulation, and detection of lightwave for communication and information processing.

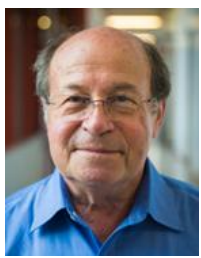

**YESHAIAHU (SHAYA) FAINMAN** (Life Fellow, IEEE) is an inaugural ASML/Cymer Chair Professor of Advanced Optical Technologies and Distinguished Professor in Electrical and Computer Engineering (ECE) at the University of California, San Diego (UCSD). He received M. Sc and Ph. D degrees from Technion-Israel Institute of Technology in 1979 and 1983, respectively. He is directing research of the Ultrafast and Nanoscale Optics group at UCSD and made significant contributions to near field optical phenomena, nanoscale science and engineering of ultra-small, sub-micrometer semiconductor light emitters and nanolasers, inhomogeneous and meta-materials, nanophotonics and Si Photonics. His current research interests are in near field optical science and technology with Si Photonics applications to information technologies and biomedical sensing. He contributed over 340 manuscripts in peer review journals and over 560 conference presentations and conference proceedings. He contributed to editorial and conference committee works of various scientific societies including IEEE, SPIE and OPTICA. He is a Fellow of OPTICA (former OSA), Fellow of the IEEE, Fellow of the SPIE, and a recipient of the Miriam and Aharon Gutvirt Prize, Lady Davis Fellowship, Brown Award, Gabor Award, Emmett N. Leith Medal, Joseph Fraunhofer Award/Robert M. Burley Prize and OPTICA Holonyak Award.